\documentclass[PRD,twocolumn,showpacs,superscriptaddress,preprintnumbers,nofootinbib,amsmath,amssymb]{revtex4}

\usepackage{graphicx}% Include figure files
\usepackage{dcolumn}% Align table columns on decimal point
\usepackage{bm}% bold math
\usepackage{color}

\def\fnl{{f_{\rm{nl}}}}
\def\fnle{{\widehat \fnl}}
\def\hnl{{f_{\rm{nl}}}}

\def\hnlb{{\widehat{\hnl}}}

\def\VEV#1{\left\langle #1 \right\rangle}

\def\fnlnull{\fnle}
\def\fnlnew{\widehat{ f_{nl}^n}}

\newcommand{\beq}{\begin{equation}}
\newcommand{\eeq}{\end{equation}}
\newcommand{\beqa}{\begin{eqnarray}}
\newcommand{\eeqa}{\end{eqnarray}}

\newcommand{\Npix}{N_{\mathrm{pix}}}

\newcommand{\lmax}{{l_{\mathrm{max}}}}

\begin{document}

\title{The Probability Distribution for Non-Gaussianity Estimators}

\author{Tristan L.~Smith}
\affiliation{Berkeley Center for Cosmological Physics, Physics
     Department, University of California, Berkeley, CA 94720} 
\author{Marc Kamionkowski}
\affiliation{California Institute of Technology, Mail Code 350-17,
Pasadena, CA 91125}
\author{Benjamin D. Wandelt}
\affiliation{UPMC Univ Paris 06, Institut dÕAstrophysique de 
Paris, 98 bis, blvd Arago, 75014 Paris, France}

\date{\today}

\begin{abstract}
One of the principle efforts in cosmic microwave background
(CMB) research is measurement of the parameter $\fnl$ that
quantifies the departure from Gaussianity in a large class of
non-minimal inflationary (and other) models.  Estimators for
$\fnl$ are composed of a sum of products of the
temperatures in three different pixels in the CMB map.  Since the 
number $\sim \Npix^2$ of terms in this sum exceeds the number
$\Npix$ of measurements, these $\sim \Npix^2$ terms cannot be
statistically independent.
Therefore, the central-limit theorem does not necessarily apply,
and the probability distribution function
(PDF) for the $\fnl$ estimator does not necessarily approach a
Gaussian distribution for $N_{\rm pix} \gg1$.  Although the variance 
of the estimators is known, the significance of a 
measurement of $\fnl$ depends on knowledge of the full shape 
of its PDF.  Here we use Monte Carlo realizations of CMB maps to
determine the PDF for two minimum-variance estimators: the standard
estimator, constructed under the null hypothesis ($\fnl=0$), and 
an improved estimator with a smaller variance for $\fnl \neq0$.  
While the PDF for the null-hypothesis 
estimator is very nearly Gaussian when 
the true value of $\fnl$
is zero, the
PDF becomes significantly non-Gaussian when $\fnl \neq 0$.   In this 
case we find that the PDF for the null-hypothesis
estimator $\fnle$ is skewed, with a long non-Gaussian tail at
$\fnle > |\fnl|$ and less probability at $\fnle < |\fnl|$ than in the
Gaussian case.  We provide an analytic fit to these PDFs.  
On the other hand, we find that the PDF for the 
improved estimator is nearly Gaussian for observationally 
allowed values of $\fnl$.
We discuss briefly the implications for
trispectrum (and other higher-order correlation) estimators.
\end{abstract}


\maketitle

\section{Introduction}

The simplest single-field slow-roll inflation models predict
that primordial perturbations should be nearly Gaussian \cite{inflation},
but with predictably small departures from Gaussianity
\cite{localmodel}.  This is often quantified through
the non-Gaussianity parameter $\fnl$ defined by
\cite{Luo:1993xx},
\begin{equation}
     \Phi = \phi + \fnl \left(\phi^2-\VEV{\phi^2}\right),
\label{fnldefinition}
\end{equation}
where $\Phi$ is the gravitational potential and $\phi$ a
Gaussian random field.  Standard single-field slow-roll
inflation predicts $\fnl \ll 1$ for the primordial field
(although nonlinear evolution of the density
field may produce $\fnl\sim 1$ at the time of recombination;
see, e.g., Ref.~\cite{Bartolo:2004if}).
However, multi-field \cite{larger} or curvaton \cite{curvaton}
models, or models with sharp features \cite{Wang:2000} or
wiggles \cite{Hannestad:2009yx} may produce larger values of
$\fnl$.  Measurement of $\fnl$ has thus become one of
the primary goals of cosmic microwave background (CMB) and
large-scale-structure (LSS) research.  Current limits from the
CMB/LSS are in the ballpark of $|\fnl|\lesssim 100$
\cite{limits,halos}.  The plot has thickened with a suggestion
\cite{yadavdetection} (not universally accepted) that WMAP data
prefers (at the $2.8\sigma$ level) $\fnl\neq0$, with a best-fit
value $\fnl\simeq35$.  The Planck satellite \cite{:2011ah} is
expected to achieve a sensitivity of $\fnl\sim5$.

In this paper, we address the following question: What is the
probability distribution function (PDF) $P(\fnle)$ for
an estimator $\fnle$ that is constructed from a CMB
map?  If the PDF departs from the Gaussian distribution that is
often assumed, then the 99.7\% confidence level (C.L.) interval for $\fnl$ may be
different than three times the standard deviation for $\fnl$.  The
interpretation of measurements thus requires knowledge of this
PDF.

The question arises as the theory
predicts not only the {\it mean} value of the estimator
$\fnle$, but it also makes a prediction for the
detailed functional form of the PDF $P(\fnle)$.  The consistency 
of a given measurement of $\fnle$ with a theoretical 
prediction for $\fnl$ depends on knowledge of the shape of $P(\fnle)$.
  Thus,
for example, we often evaluate or forecast the standard error
$\sigma_{\fnl}$ with which a given measurement will recover the
true value of $\fnl$ and then simply assume that the error is
Gaussian.  If so, then with $\sigma_{\fnl}=10$, for example, a
measurement of $\fnle=30$ would represent a $3\sigma$ departure
from $\fnl=0$ and a measurement $\fnle=0$ would represent a
$3\sigma$ departure fom $\fnl=30$.  However, if the PDF depends
on the true value $\fnl$, and if that distribution is
non-Gaussian, then it may be that a measurement $\fnle=30$ could
be easily consistent with a true value $\fnl=0$, while a
measurement $\fnle=0$ could be inconsistent with $\fnl=30$ with
a confidence greater than ``$3\sigma$.''   We will see
below that something like this actually occurs with measurements
of $\fnl$.

This question is particularly important for measurements of
non-Gaussianity (as opposed, for example, for the CMB power
spectrum), because $\fnle$ is a sum over products of three
temperature measurements (unlike the power spectrum, which sums
over squares of temperature measurements).  Suppose the
temperature is measured in $\Npix$ pixels.  There are then $\sim
\Npix^2$ terms in the $\fnl$ estimator (after restrictions
imposed by statistical isotropy).  While these terms may have
zero covariance, they are not statistically independent; there
is no way to construct $\Npix^2$ statistically independent
quantities from $\Npix$ measurements! The conditions required
for the validity of the central-limit theorem are therefore not
met, and $P(\fnle)$ will not necessarily approach a Gaussian in the 
$\Npix \gg 1$ limit.

The PDF can be obtained from Monte Carlo simulations,
but the simulations are very computationally intensive
(e.g., Ref.~\cite{Elsner:2009md}).  The number of Monte Carlo realizations is thus usually limited
to the number, $\lesssim1000$, required to determine a 99.7\%
C.L.\ detection or sometimes even fewer if it is just the variance
that is being estimated.  Although with only 1000 realizations Fig.~8 in 
Ref.~\cite{Elsner:2009md} shows hints of a non-Gaussian PDF, 
simulations done up until now do not include enough realizations
to precisely map the functional form of $P(\fnle)$.  
The number of realizations required to map ultimately the
$4\sigma$, $5\sigma$, etc.\ ranges will be prohibitive, 
especially since the simulations will need to be re-run
repeatedly to determine how the error ranges depend on
cosmological parameters, instrument-noise properties, scanning
strategies, etc., and they then must be run for multiple
theoretical values $\fnl$.

Work along these lines was begun in
Ref.~\cite{Creminelli:2006gc}, wherein it was shown that the
variance of the distribution $P(\fnle)$ may have a strong
dependence on the true underlying value of $\fnl$.  More
precisely, they evaluated the variance of the estimator designed
to have the minimum variance under the null hypothesis $\fnl=0$
(which we refer to frequently below as the ``null-hypothesis
minimum-variance'' estimator, or NHMV estimator), and showed
that the variance of this NHMV estimator increases as $\fnl^2$ 
increases.  They then constructed an alternative estimator
$\fnlnew$, which we call the CSZ estimator\footnote{We note that the CSZ estimator, 
which is defined under the Sachs-Wolfe limit, 
has yet to be generalized so that it can be applied to actual data.  On the 
other hand a  
Bayesian approach, discussed in Ref.~\cite{Elsner:2010hb}, allows for an $\fnl$ inference 
that saturates the Cramer-Rao bound even in the presence of non-Gaussianity.}, which has a PDF with 
a variance that saturates the Cramer-Rao bound up to corrections 
of order $\fnl^2$.  
Still, as we have argued above,
the consistency of a hypothesis with a measurement requires full
knowledge of the PDF of whatever estimator is used in the
analysis.

To address these questions, we calculate the PDF for an
ideal (no-noise) map to understand the irreducible PDF introduced
simply by cosmic variance under the Sachs-Wolfe approximation 
and on a flat sky.  We hope that lessons learned about
$P(\fnle)$ in this ideal case may help interpret and understand
current/forthcoming results and assess the validity of
full-experiment simulations.  

We calculate these PDFs by using
Monte Carlo realizations of numerous no-noise flat-sky CMB
maps.  The first order of business with a map will be to determine
whether a given map is consistent or inconsistent with the null
hypothesis $\fnl=0$.  Therefore, we first calculate the PDF that
arises if $\fnl$ does indeed vanish, for the
NHMV estimator $\fnle$,
and we also calculate the PDF that
arises if the true value of $\fnl$ is nonzero.  We provide an analytic 
fit for these PDFs in Eq.~(\ref{eq:fit}).  If the evidence
from such a measurement were to show that $\fnl$ is nonzero,
then the next step would be to apply the CSZ estimator $\fnlnew$ for $\fnl\neq0$
\cite{Creminelli:2006gc} to obtain a more precise value for
$\fnl$ or to test consistency of the data with a specific
nonzero value of $\fnl$.  We therefore follow by calculating the
PDF for these improved non-null-hypothesis estimators.

We find that, besides having a variance that 
increases with $\fnl^2$, 
the PDF of the NHMV can have a significantly non-Gaussian shape
 when $\fnl \neq0$ with a long non-Gaussian tail for
$\fnle>|\fnl|$ and less probability at $\fnle < |\fnl|$ than in the
Gaussian case.  As an example, taking $\fnl = 100$ 
for an experiment which measures 
multipoles out to $l_{\rm max} = 3000$ (such as Planck) and \emph{assuming
a Gaussian PDF} for the NHMV this experiment measures $74 \leq \fnle \leq 148$ 
at the 99.7\% C.L.; the actual PDF shows that this experiment 
measures $68 \leq \fnle \leq 143$ at the 99.7\% C.L.\
Applying the CSZ estimator to the data 
we find it has a PDF which is well approximated by a 
Gaussian with 
$\fnle = 100 \pm 12.5$ at 99.7\% C.L.\

This paper is organized as follows.  In Sec.~\ref{sec:approxs} we
construct the standard minimum-variance estimator $\fnle$ under
the null hypothesis $\fnl=0$ and discuss why the PDF for this
estimator is not necessarily Gaussian, even in the limit of
a large number of pixels.  In  Sec.~\ref{sec:null} we use
Monte Carlo calculations to evaluate the PDF $P(\fnle$) for this
estimator if the null hypothesis is indeed valid, i.e., if $\fnl$ is
indeed zero.  We find that the PDF in this $\fnl=0$ case is well
approximated by a Gaussian, for $\Npix \gg 1$, even though the central-limit
theorem does not apply.  In Sec.~\ref{sec:non_zero_fNL}, we calculate
the PDF assuming that the null hypothesis is not valid, i.e., if
$\fnl \neq 0$.  We find the PDFs in this case can be highly
non-Gaussian, skewed to large $|\fnle|$, with long large-$\fnle$
non-Gaussian tails and less likelihood at $\fnle\leq|\fnl|$ relative
to the Gaussian distribution of the same variance.  We provide
fitting formulas for the PDF as a function of the estimator
$\fnle$, the true value of $\fnl$, and the maximum multipole
moment $\lmax$ of the map.  In Sec.~\ref{sec:improved} we
discuss the PDF of the CSZ estimator.  We show
that this estimator is well approximated by a Gaussian for
values of $\fnl$ still allowed by observations.  In
Sec.~\ref{sec:discussion} we summarize and discuss
some possible implications of the work for other bispectra and
also for the trispectrum and other higher-order statistics.  An
Appendix discusses the computational techniques we used in order to 
perform our Monte Carlo simulations.

\section{Non-Gaussianity estimators \label{sec:approxs}}

\subsection{Formalism} 

We assume a flat sky to avoid the complications (e.g., spherical
harmonics, Clebsch-Gordan coefficients, Wigner 3$j$ and 6$j$
symbols, etc.) associated with a spherical sky, and we further
assume the Sachs-Wolfe limit.  We denote the fractional
temperature perturbation at position $\vec\theta$ on a flat sky
by $T(\vec\theta)$ and refer to it hereafter simply as the
temperature.

The field $T(\vec\theta)$ has a power spectrum
$C_l$ given by
\begin{equation}
     \VEV{T_{\vec l_1} T_{\vec l_2}} = \Omega \delta_{\vec
     l_1+\vec l_2,0} C_l,
\label{eqn:powerspectrum}
\end{equation}
where $\Omega=4\pi f_{\mathrm{sky}}$ is the survey area (in
steradian), 
\begin{equation}
     T_{\vec l} = \int\, d^2\vec \theta\, e^{-i\vec l\cdot
     \vec \theta} T(\vec\theta) \simeq
     \frac{\Omega}{N_{\mathrm{pix}}} \sum_{\vec\theta} e^{-i\vec l\cdot
     \vec \theta} T(\vec\theta),
\end{equation}
is the Fourier transform of $T(\vec\theta)$, and $\delta_{\vec
l_1+\vec l_2,0}$ is a Kronecker delta that sets $\vec l_1 =
-\vec l_2$. The power spectrum
for $T(\vec\theta)$ is given by
\begin{equation}
     C_l = \frac{2 \pi A}{l^2},
     \label{eq:Cl}
\end{equation}
where the amplitude, $A \simeq 10^{-10}$.
The bispectrum $B(l_1,l_2,l_3)$ is  defined by 
\begin{equation}
     \VEV{T_{\vec l_1} T_{\vec l_2} T_{\vec l_3}} = \Omega
     \delta_{\vec l_1 +\vec l_2 +\vec l_3,0} B(l_1,l_2,l_3).
\label{eqn:bispectrum}
\end{equation}
The Kronecker delta insures that the bispectrum is defined only
for $\vec l_1 +\vec l_2+\vec l_3=0$; i.e., only for triangles in
Fourier space.  Statistical isotropy then dictates that the
bispectrum depends only on the magnitudes $l_1$, $l_2$, $l_3$ of
the three sides of this Fourier triangle.  

\subsection{The null-hypothesis minimum-variance estimator}
\label{sec:estimators}

We now review how to construct the minimum-variance estimator 
for $\hnl$ under the null hypothesis.  This
is the quantity that one would first determine from the data to
check for consistency of the measurement with the null
hypothesis $\fnl=0$.  

From Eq.~(\ref{eqn:bispectrum}), each triangle $\vec l_1 +\vec
l_2 +\vec l_3 =0$ gives an estimator,
\begin{equation}
     (\hnlb)_{123} = \frac{ T_{\vec l_1} T_{\vec l_2} T_{\vec
     l_3} }{\Omega B(l_1,l_2,l_3)/\hnl},
\label{eqn:onetriangle}
\end{equation}
and under the null hypothesis this has a variance proportional to
\begin{equation}
     \frac{\Omega^3 C_{l_1} C_{l_2} C_{l_3}}{\left[ \Omega
     B(l_1,l_2,l_3)/\hnl \right]^2}.
\label{eqn:singlevariance}
\end{equation}
The null-hypothesis minimum-variance estimator is constructed by
adding all of these estimators with inverse-variance weighting.
It is \cite{Babich:2004yc,Kamionkowski:2010me}
\begin{equation}
     \fnlnull \equiv \sigma_{\fnl}^{2} \sum_{\vec l _1+ \vec l_2+
     \vec l_3=0} \frac{ T_{\vec l_1} T_{\vec
     l_2} T_{\vec l_3} B(l_1,l_2,l_3)/\hnl}{ 6\Omega^2 
     C_{l_1}C_{l_2}C_{l_3}},
\label{eqn:biestimator}
\end{equation}
and it has inverse variance,
\begin{equation}
     \sigma_{\fnl}^{-2} = \sum_{\vec l _1+ \vec l_2+ \vec l_3=0} \frac{
     \left[ B(l_1,l_2,l_3)/\hnl \right]^2}{6 \Omega
     C_{l_1}C_{l_2}C_{l_3}}.
\label{eqn:binoise}
\end{equation}

\subsection{Non-gaussianity of the PDF}

If the number of pixels in the CMB map is $\Npix$, then there are
also $\Npix$ statistically independent $T_{\vec l}$.  But there
are a much larger number, $\propto N_{\rm pix}^2$, of triplets
$T_{\vec l_1} T_{\vec l_2} T_{\vec l_3}$, included in the
estimator [cf., Eq.~(\ref{eqn:biestimator})],
and so the number of individual ``data points'' (i.e., triplets)
used in the minimum-variance estimator scales like
$N_{\rm pix}^2 \gg N_{\rm pix}$.  Since the number of
terms included in the estimator is greater than the number of
independently measured data points the standard central-limit
theorem does not apply.  Thus, we cannot assume that the PDF of the
estimator will approach a Gaussian in the $N_{\rm pix} \to
\infty$ limit.  

This contrasts with the estimator $\widehat C_l \propto \sum
|T_{\vec l}|^2$ of the power spectrum $C_l$.  While the PDF for
$\widehat C_l$ is not necessarily Gaussian (it has a
$\chi_{2l+1}^2$ distribution), it is the sum of the squares of
{\it statistically independent} quantities.  The central-limit
theorem therefore applies, and the distribution for $\widehat
C_l$ does indeed approach a Gaussian for large $l$.  The
problems we address here for $\fnl$ estimators parallel those
discussed in the literature for the quadrupole moment $C_2$, as
the distribution for quadrupole-moment estimators will be highly
non-Gaussian and will also depend on the underlying theory 
(see, e.g., Ref.~\cite{Efstathiou:2003tv}).

\section{The PDF of $\fnlnull$ for the local model}

We now restrict our attention to a family of non-Gaussian models
in which the temperature $T(\vec{\theta})$ has a non-Gaussian
component; i.e.,
\begin{equation}
     T(\vec{\theta}) = t(\vec{\theta}) + 3 \fnl \left\{ [t(\vec{\theta})]^2
     -\VEV{[t(\vec{\theta})]^2} \right\},
\label{eqn:localmodel}
\end{equation}
where $t(\theta)$ is a Gaussian random field with a power spectrum $C_l$ 
given in Eq.~(\ref{eq:Cl}). 
 To zero-th order in $\fnl$, the power spectrum and
correlation function for $T(\vec \theta)$ are the same as those for
$t(\vec \theta)$.  Note that $T(\vec \theta)$ is, strictly speaking, the
temperature fluctuation, so $\VEV{T(\vec \theta)}=0=T_{\vec l=0}$.
The bispectrum for this model is
\begin{equation}
     B(l_1, l_2, l_3) = 6 \fnl(C_{l_1} C_{l_2} +C_{l_1}C_{l_3}+C_{l_2} C_{l_3}).
     \label{eq:localModBi}
\end{equation}  

The temperature Fourier coefficients can be written $T_{\vec
l}=t_{\vec l}+\fnl \delta t^2_{\vec l}$ with
\begin{eqnarray}
     \delta t^2_{\vec l} &\equiv & \frac{3}{\Omega} \sum_{\vec
     l'} t_{\vec l - \vec l'} t_{\vec l'}.
     \label{eq:nonGauss}
\end{eqnarray}
Formally, the sum goes from $0 < |\vec l'| \leq \infty$, but
for a finite-resolution map, the sum is truncated at some $l_{\rm
max}$ such that the number of Fourier modes equals the number of
data points.

We now proceed to evaluate $P(\fnle;\fnl,\lmax)$, the PDF that arises if
the true value is $\fnl$ for the NHMV estimator $\fnle$ and for
a map with $\lmax$.  To do so, we generated large numbers
of Monte Carlo realizations of maps according to
Eq.~(\ref{eq:nonGauss}), for some assumed value of $\fnl$, and
then applied the  estimator in Eq.~(\ref{eqn:biestimator}) to
these maps.  Each map is simulated in harmonic space from
$l_{\rm min} = 2$ up to a maximum multipole $l_{\rm max}$.  In
order to produce a large number of realizations we re-expressed
the generation of maps and implementation of the
estimator in terms of fast Fourier transforms as discussed in
Appendix A.

\subsection{The PDF of the null hypothesis minimum-variance estimator with $\fnl = 0$}
\label{sec:null}

\begin{figure}[htbp]
\resizebox{!}{6.25cm}{\includegraphics{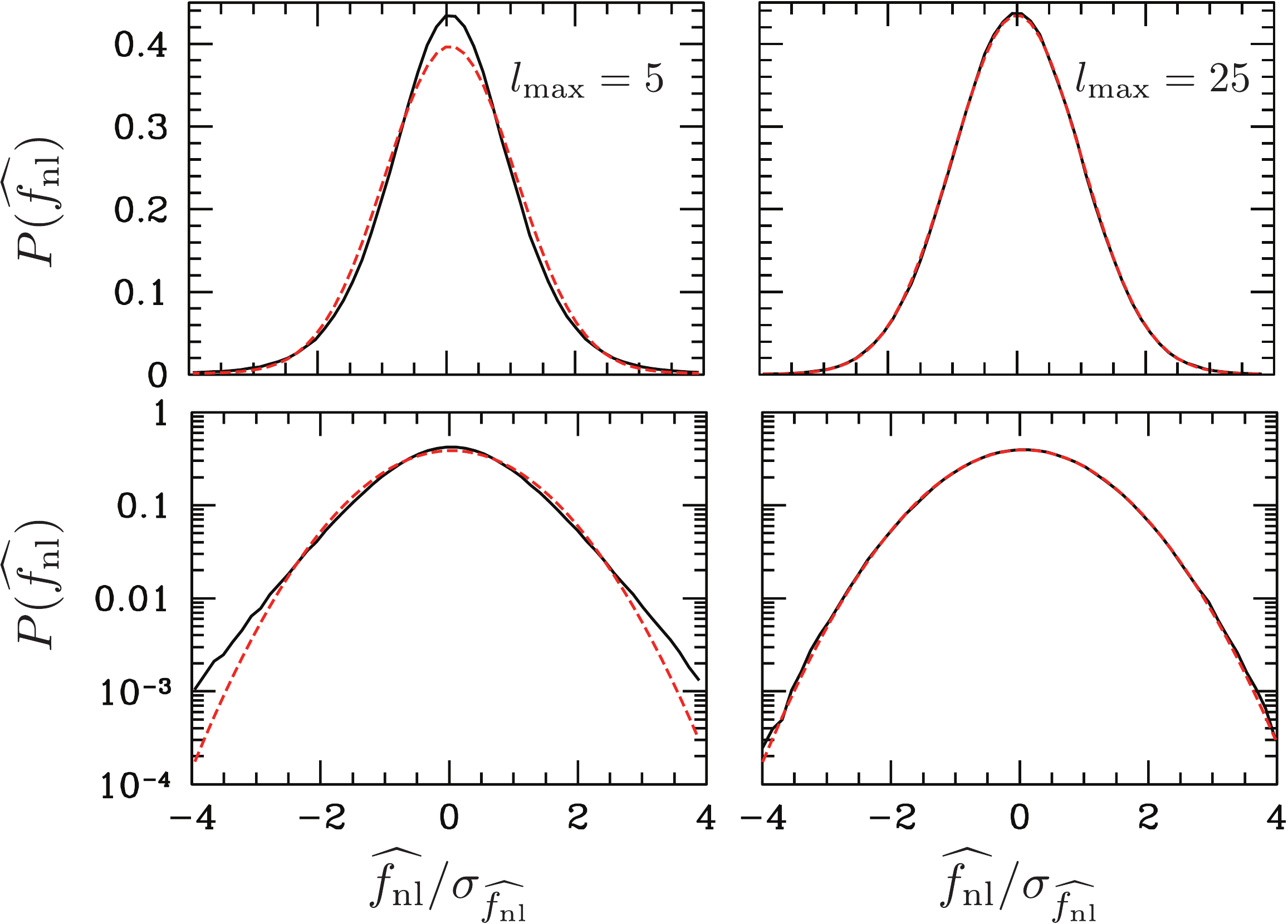}}
\caption{
     Numerical evaluations of $P(\fnlnull; \fnl=0, l_{\rm max})$.  The left
     (right) two panels show the PDF for $l_{\rm max} = 5$ and
     $l_{\rm max} = 25$ for $10^6$ realizations for a scale-invariant power
     spectrum.   In all
     panels the PDF has been normalized to have a unit variance,
     and the corresponding Gaussian PDF (with the same variance)
     is shown as the red
     dashed curve.  As $l_{\rm max}$ gets
     larger, the PDF tends towards a Gaussian.  This is not
     guaranteed by the central-limit theorem
     since the majority of the terms that appear in the
     estimator are \emph{not} statistically independent.}
\label{fig:nullPDF}
\end{figure}

First we consider the shape of $P(\fnlnull; \fnl=0, l_{\rm
max})$, the PDF for the NHMV estimator in
Eq.~(\ref{eqn:biestimator}) applied to a purely Gaussian
($\fnl=0$) map.  To do this we generated  $10^6$ Gaussian
realizations and applied the estimator in
Eq.~(\ref{eqn:biestimator}) to  generate a histogram of values
of $\fnlnull$.  From this histogram we determined $P(\fnlnull;
\fnl=0, l_{\rm max})$ out to four times the root-variance, as
shown in Fig~\ref{fig:nullPDF}.
 
First we note that our simulations verify that the variance of
the distribution for the null case is well approximated by the analytic
expression \cite{Babich:2004yc,Kamionkowski:2010me},
\begin{eqnarray}
     \sigma_{\fnl}^{2} &\approx& \frac{1}{8A  l_{\rm max}^2
     \ln(l_{\rm max})}.
\end{eqnarray}

Additionally our simulations show that out to at least four times the
root-variance, the PDF $P(\fnlnull; \fnl=0, l_{\rm max})$ is
well approximated by a Gaussian for $l_{\rm max} \gtrsim 25$,
even though the conditions for the central-limit theorem to
apply are not satisfied.  Therefore, a measurement of $\fnle$
that differed from 0
at more than three times the root-variance would indeed
constitute a `99.7\% confidence level' inconsistency with the
$\fnl = 0$ hypothesis.

\subsection{The PDF of the null hypothesis minimum-variance
estimator with $\fnl \neq 0$}
\label{sec:non_zero_fNL}

We now consider the form of $P(\fnlnull; \fnl, l_{\rm max})$
when $f_{\rm nl} \neq 0$, the PDF for the null-hypothesis
minimum-variance estimator if the null hypothesis is in fact not
valid.  In this case, the non-Gaussian statistics of the
$T_{\vec l}$s impart some non-Gaussianity to the $\fnlnull$
PDF.

\begin{figure}[htbp]
\resizebox{!}{11cm}{\includegraphics{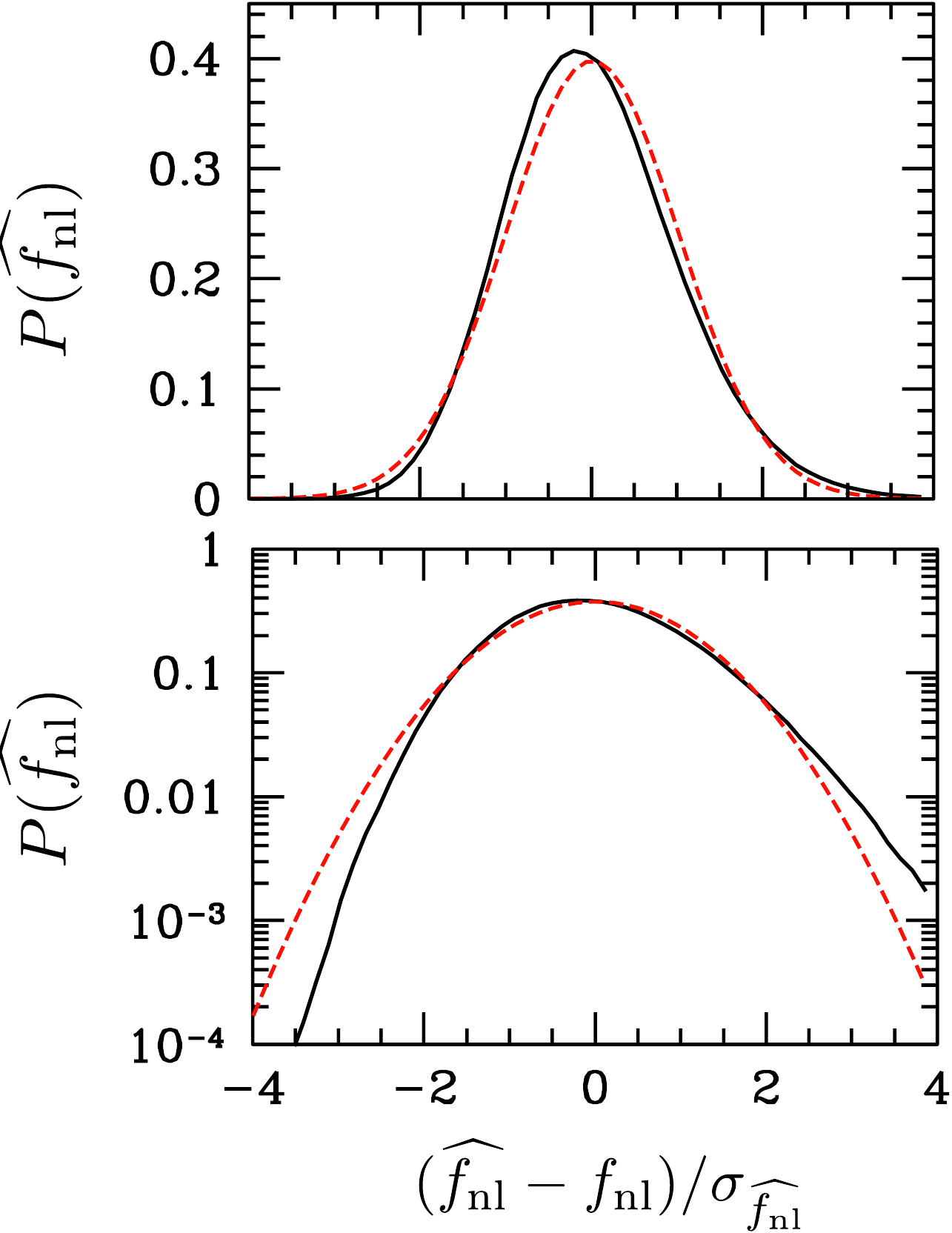}}
\caption{The PDF $P(\fnle)$ when $\fnle = 1500 $ using the estimator in
     Eq.~(\protect\ref{eqn:biestimator}) with $l_{\rm max} = 25$.  
     The upper (lower) panel shows the PDF on a linear (log) scale.  
     We can see that the PDF is significantly non-Gaussian with an exponential 
     drop-off to the left of mean and a power-law to the right.  
     We provide a fitting formula
     for $P(\fnlnull; \fnl,l_{\rm max})$ in the text. }
\label{fig:non-GaussPDF}
\end{figure}

In Fig.~\ref{fig:non-GaussPDF} 
we show $P(\fnlnull; \fnl, l_{\rm max})$ calculated using 
$10^6$ realizations with $\fnl = 1500$ and $l_{\rm max} = 25$.  Clearly 
the PDF in this case is highly non-Gaussian. 

Non-Gaussianity of $P(\fnlnull;\fnl,l_{\rm max})$ for a central value
$\fnl\neq0$ may be significant for the interpretation of data.
Suppose, for example, that a CMB measurement returns $\fnlnull =0$
with a root-variance $\sigma_{\fnl} =40$.  If the
PDF was assumed to be Gaussian the measurement $\fnlnull=0$ 
would rule out $\fnl=100$ at the $2.5\sigma$ level, 
but given the
asymmetric PDF of Fig.~\ref{fig:non-GaussPDF} it may rule out $\fnl =100$ at 
a much higher significance. 

In order to better understand the origin of the non-Gaussian
PDF, it is useful to expand the minimum-variance estimator in
Eq.~(\ref{eqn:biestimator}) to linear order in $\fnl$
\cite{Creminelli:2006gc}:
\begin{equation}
     \fnlnull \approx{E}_0 + \fnl{E}_1 + \cdots,
\end{equation}
where 
\begin{eqnarray}
     {E}_0 &=& \sigma_{\fnl}^2 \sum_{\vec l_1 +\vec l_2+ \vec l_3=0} 
     \frac{t_{\vec l_1}t_{\vec l_2}t_{\vec l_3}}{6 \Omega^2 \fnl
     C_{l_1} C_{l_2} C_{l_3} }B(l_1,l_2,l_3) \\
     {E}_1 &=& \sigma_{\fnl}^2 \sum_{\vec l_1+ \vec
     l_2+ \vec l_3=0} \frac{\delta t^2_{\vec l_1}t_{\vec
     l_2}t_{\vec l_3}}{2  \Omega^3 C_{l_1} C_{l_2} C_{l_3}
     }B(l_1,l_2,l_3).
\label{eq:E1}
\end{eqnarray}
Since ${E}_0 \sim t^3$ and ${E}_1 \sim t^4$, it is clear that
$\langle{E}_0 \rangle = 0$ and $\langle{E}_0{E}_1 \rangle = 0$, and
the normalization guarantees that $\VEV{{E}_1} =1$.
Furthermore, since we have already established that
$P(\fnlnull)$ approaches a Gaussian in the large $l_{\rm max}$
limit if $\fnl =0$, we know that, to leading order, the
non-Gaussian shape of $P(\fnlnull;\fnl,l_{\rm max})$ for $\fnl
\neq 0$ is being generated by ${E}_1$.  

\begin{figure}[htbp]
\resizebox{!}{8.5cm}{\includegraphics{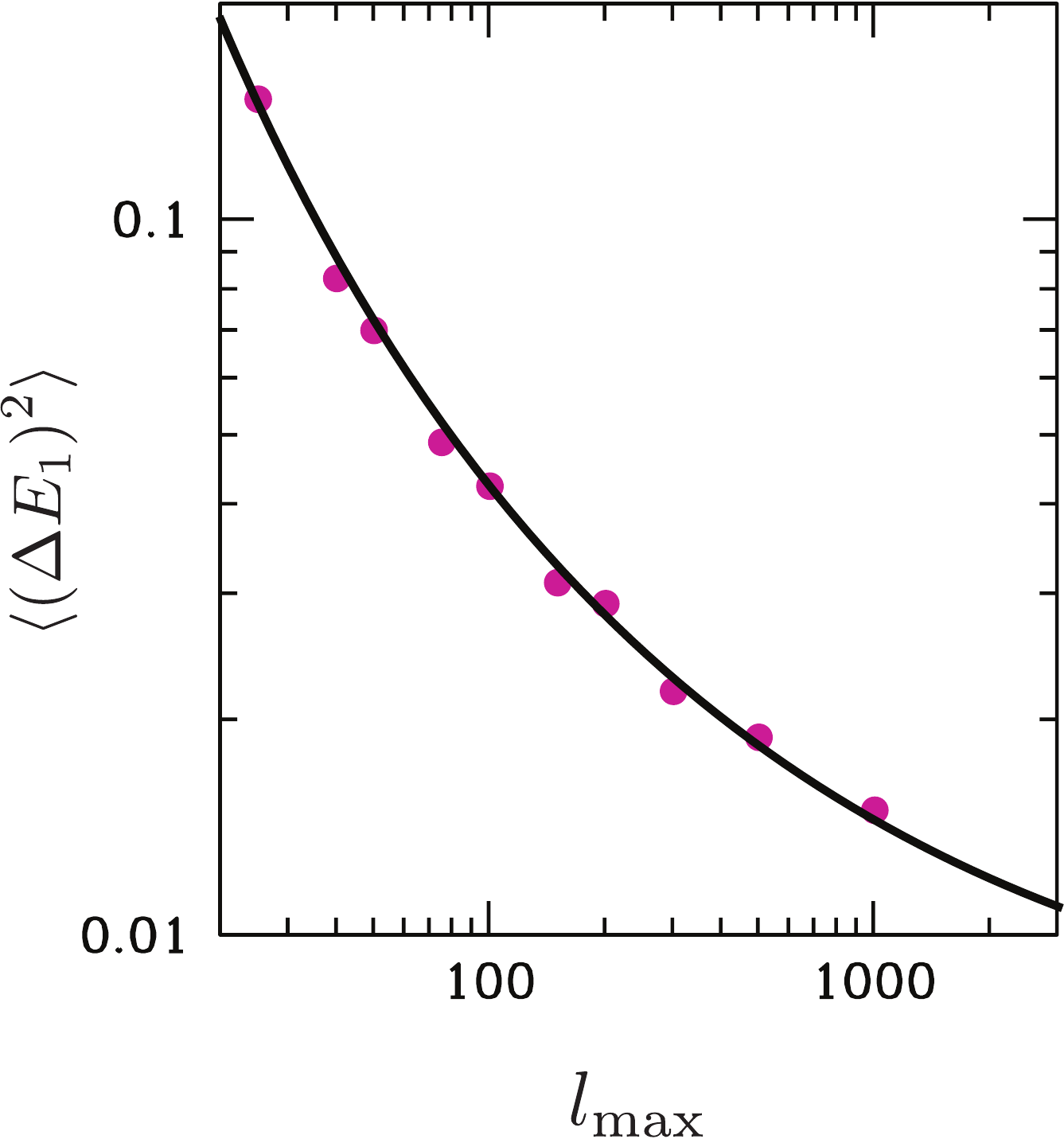}}
\caption{The dependence of $\VEV{(\Delta{E}_1)^2}$ on $l_{\rm
     max}$.  The points correspond to the results of our Monte
     Carlo simulations for 1000 realizations at different values
     of $l_{\rm max}$.  The solid curve shows the analytic
     calculation of the variance presented in Appendix B which
     is well-fit by the function $\VEV{(\Delta{E}_1)^2} = [14.0
     (l_{\rm max})^{0.433}]/[\ln^{5.1}(l_{\rm max})] \approx
     4.5\ln^{-3}(l_{\rm max})$.} 
\label{fig:varE1}
\end{figure}

Some of the statistics associated with ${E}_1$ have already been explored in 
Ref.~\cite{Creminelli:2006gc}.  There it is noted that 
the variance of $\fnlnull$ is dominated by ${E}_1$ in the high $S/N$ limit 
leading to a slower scaling of the $S/N$ than the $l_{\rm max}^{-2}\ln^{-1}(l_{\rm max})$ scaling expected 
if the estimator saturated the Cramer-Rao bound \cite{Creminelli:2006gc}.
We explored the same limit using our Monte Carlo realizations,
as shown in Fig.~\ref{fig:varE1}, and find the same qualitative
trend but with a different dependence on 
$l_{\rm max}$.  Ref.~\cite{Creminelli:2006gc} found 
$\langle (\Delta{E}_1)^2 \rangle \propto \ln^{-2}(l_{\rm max})$ whereas our 
simulations show $\langle (\Delta{E}_1)^2 \rangle \propto
\ln^{-3}(l_{\rm max})$.  We have checked the scaling found with
our simulations by computing the variance analytically, as we
further discuss in Appendix B.  Fig.~\ref{fig:varE1} shows the
agreement between our analytic calculation (solid curve) and
simulations (data points).

\begin{figure}[htbp]
\resizebox{!}{8cm}{\includegraphics{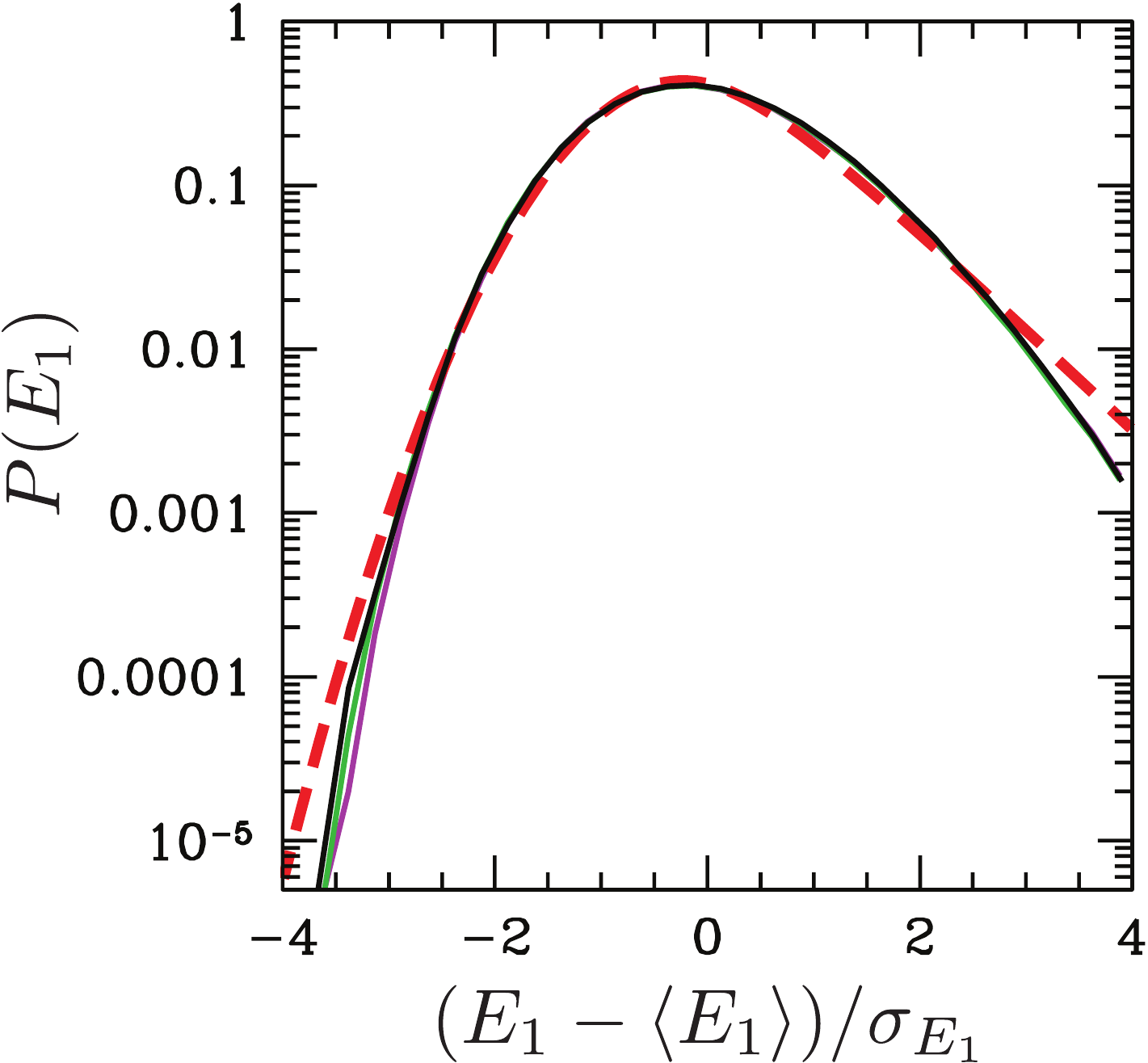}}
\caption{The PDF of ${E}_1$ calculated using $10^6$
     realizations. The thin solid curves correspond to
     $P({E}_1)$ with $l_{\rm max} = 25$ (green), $l_{\rm max} =
     50$ (purple), $l_{\rm max} = 100$ (black).  Since the
     functional form of the PDF for each choice of $l_{\rm max}$
     is nearly identical, we conclude that $P({E}_1)$ approaches
     an asymptotic functional form in the large-$l_{\rm max}$
     limit.  The thick red dashed curve corresponds to a fit to
     $P({E}_1)$, accurate to $\sim$10\% out to 3 times the root variance, using the fitting formula in
     Eq.~(\ref{eq:fitting}) with parameter values $ x_p = -0.22$, $\sigma = 0.80$, and $c = 0.91$.  }
\label{fig:PDFE1}
\end{figure}
Our simulations allow us to generate the full PDF for $E_1$, not
just the variance.  
Fig.~\ref{fig:PDFE1} shows this PDF for
various choices of $l_{\rm max}$ (thin solid lines).  An important
conclusion from Fig.~\ref{fig:PDFE1} is that the shape of the
PDF approaches a universal form in the $l_{\rm max} \gg 1$
limit.  
We provide a fit to the PDF (thick red dashed line), accurate to $\sim$10\% (40\%) out to three (four) times the 
root variance, using the fitting formula
\begin{eqnarray}
   && \log[ F(x)] = N- \label{eq:fitting}\\
 &&    \begin{cases}
      -(x-x_p)^2/(2 \sigma^2), & x \leq x_p\\
      -\frac{c}{\sigma^2}\left(\sqrt{(x-x_p) ^2 + c^2}-c\right), & x > x_p,
\end{cases}\nonumber
\end{eqnarray}
where $N \equiv \sqrt{2/\pi} \sigma + c \exp[c^2/\sigma^2] K_1(c^2/\sigma^2)$ and $K_1(x)$ is a 
modified Bessel function of the first kind, 
$c$ quantifies the non-Gaussianity of the distribution (and approaches a 
Gaussian in the $c \rightarrow \infty$ limit) and $x_p$ is the value of $(E_1-\VEV{E_1})/\sigma_{E_1} $ at the peak of the
distribution. 
The red curve in Fig.~\ref{fig:PDFE1} shows Eq.~(\ref{eq:fitting}) with 
parameter values $ x_p
= -0.22$, $\sigma = 0.80$, and $c = 0.91$.  

We are now in a position to write down a semi-analytic
 expression for $P(\fnlnull;\fnl,l_{\rm max})$, accurate to $\sim$10\% 
 (40\%) out to three (four) times the 
root variance,
 as a function of 
$\fnl$ and $l_{\rm max}$.  Letting $\sigma_0$ and $\sigma_1$
denote the standard deviations of  
the distributions for ${E}_0$ and ${E}_1$ respectively we have
\begin{eqnarray}
\sigma_0^2 &\approx& \frac{1}{8 A l_{\rm max}^2 \ln(l_{\rm max})}, \\
\sigma_1^2 &\approx& \frac{9\fnl^2}{2\ln^3(l_{\rm max})}.
\end{eqnarray}
A good approximation to the PDF of $\fnlnull$ is provided
by the convolution of the PDF of ${E}_0$ and $\fnl {E}_1$:
\begin{eqnarray}
     P(\fnlnull;\fnl,l_{\rm max}) &\approx&
     \frac{4}{9\sqrt{2\pi} \sigma_0 \sigma_1}\\ 
   &\times&  \int_{-\infty}^{\infty} G_0(\fnlnull-x)
     F([x-\fnl]/\sigma_1) dx, \nonumber
\label{eq:approxPDF}
\end{eqnarray}
where $G_0(x)$ is a Gaussian with zero mean and standard deviation $\sigma_0$ and 
$F([x-\fnl]/\sigma_1)$ is given by Eq.~(\ref{eq:fitting}) with $
x_p = -0.22$, $\sigma = 0.80$, and $c = 0.91$.

To obtain an analytic expression for the PDF we can approximate the convolution 
in Eq.~(\ref{eq:approxPDF}) to write
\begin{widetext}
\begin{eqnarray}
&&P(\fnlnull;\fnl,l_{\rm max}) \approx \frac{2}{9}\exp\left[ - \frac{X^2}
{2 (\sigma_0^2 + \sigma_1^2 \sigma^2)}\right] \sqrt{\frac{1}{ \sigma_1^2 (\sigma_0^2 + \sigma_1^2 \sigma^2)}}
\Bigg\{ \sigma_1 \sigma \left(1+{\rm erf}\left[\frac{\sigma_0^2 + \sigma_1\sigma^2( X+ \sigma_1)}{\sigma_0 \sigma \sqrt{2(\sigma_0^2 + \sigma_1^2 \sigma^2)}}\right]\right)\label{eq:fit} \\
&+& \sqrt{\sigma_0^2 + \sigma_1^2 \sigma^2}\left(1-{\rm erf}\left[\frac{c \sigma_0^2 + \sigma_1\sigma^2(X + \sigma_1 )}{\sqrt{2} \sigma_0 \sigma_1 \sigma^2}\right]\right)\exp\left[\frac{1}{2} \left(\frac{c^2 \sigma_0^2}{\sigma_1^2 \sigma^4} + \frac{2 c [X+ \sigma_1c^2 ]}{\sigma_1 \sigma^2} + \frac{X^2}{\sigma_0^2 +
\sigma_1^2 \sigma^2}\right)\right]\Bigg\},\nonumber
\end{eqnarray}
\end{widetext}
where $X \equiv \fnl+x_p \sigma_1- \widehat{\fnl}$.

\begin{figure}[htbp]
\resizebox{!}{7.3cm}{\includegraphics{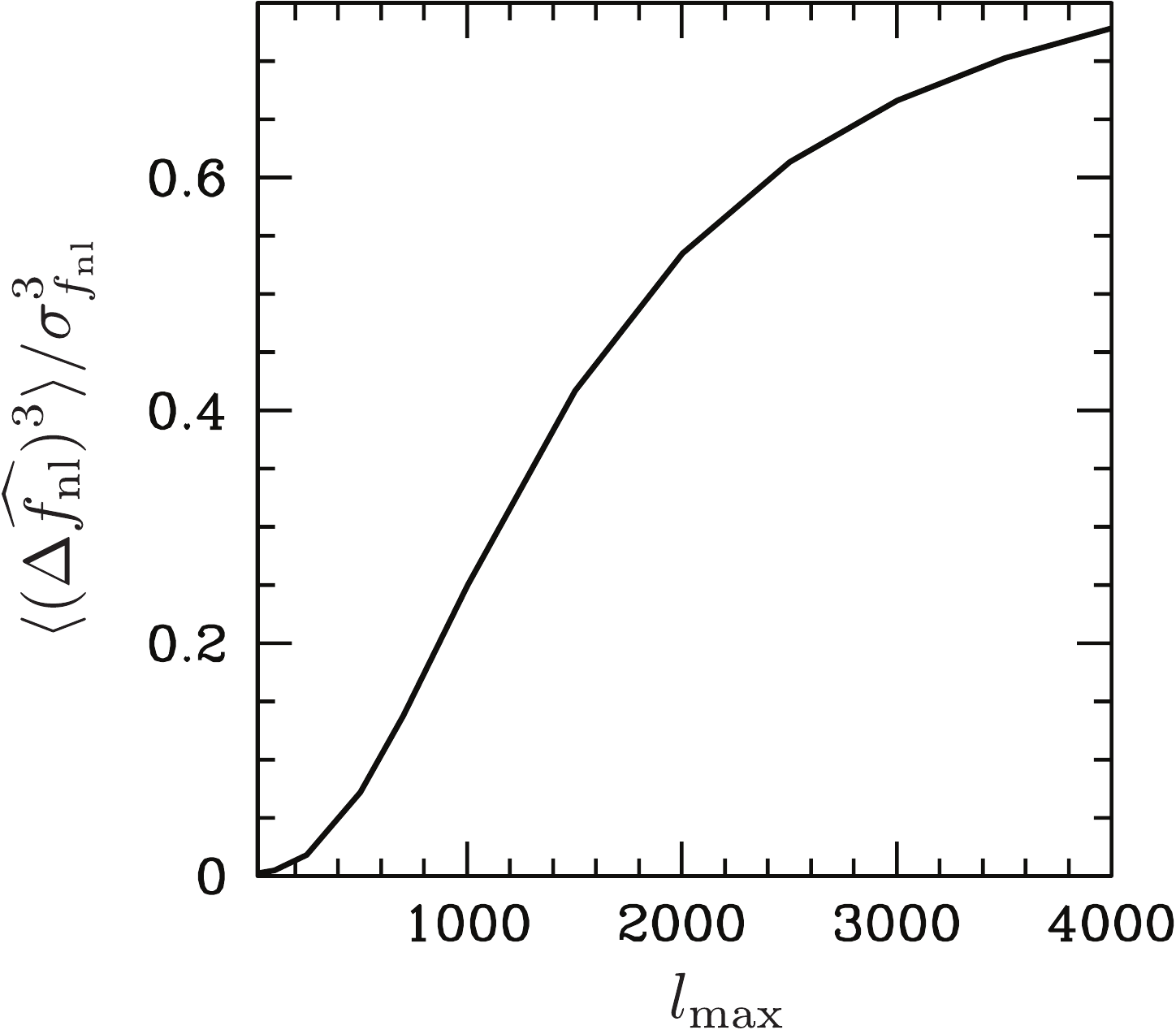}}
\caption{The skewness, $\langle (\Delta \fnle)^3 \rangle$, as a fraction of the 
variance of $P(\fnlnull;\fnl,l_{\rm
max})$ as a function of $l_{\rm max}$ for $\fnl = 100$.  We provide an analytic 
fitting formula in Eqs.~(\ref{eq:skew}) and (\ref{eq:var}) as a function of $\fnl$ and $l_{\rm max}$.  }
\label{fig:skew}
\end{figure}

Another useful way of quantifying the non-Gaussian shape of $P(\fnlnull;\fnl,l_{\rm
max})$ is to measure its skewness, $\langle (\Delta \fnle)^3 \rangle$, as a function of $\fnl$ and $l_{\rm max}$.  We 
show this in Fig.~\ref{fig:skew} for $\fnl = 100$.  An analytic fit to the skewness is given by 
\begin{eqnarray}
&&\frac{\langle (\Delta \fnle)^3 \rangle}{\sigma_{\fnl}^3} = \left(\frac{\fnl}{100}\right)^3 \label{eq:skew} \\
&\times&\left( \frac{1}{1+3.7 \exp\left[-
(l_{\rm max}-5.1)/740\right]} - 0.26 \right),\nonumber
\end{eqnarray}
with the variance of the distribution, $\sigma_{\fnl}^2$, given by
\begin{equation}
     \sigma_{\fnl}^2 \approx \frac{1}{8 A l_{\rm max}^2
     \ln(l_{\rm max})}\left[1 + \frac{36 A \fnl^2 l_{\rm
     max}^2}{\ln^2(l_{\rm max})}\right].
     \label{eq:var}
\end{equation}

Finally, we note that the shape of $P(\fnlnull;\fnl,l_{\rm
max})$ departs significantly from a Gaussian when $\sigma_0 \simeq  \sigma_1$.  This
occurs when 
\begin{equation}
     \fnl A^{1/2} \gtrsim \frac{\ln(l_{\rm max})}{6 l_{\rm max}}.
\end{equation}
Therefore, for the Planck satellite
(i.e., $l_{\rm max} = 3000$) the non-Gaussian 
features of $P(\fnlnull;\fnl,l_{\rm max})$ for the NHMV estimator 
are significant if $\fnl \gtrsim
\mathcal{O}(10)$.  Thus, given that Planck is expected to measure 
$\fnl$ with a variance $\sigma \approx 5$, 
these PDFs may need to be taken into account to
assign a precise confidence region with Planck data.

\section{The PDF of an improved estimator when $\fnl \neq 0$ }
\label{sec:improved}

As we saw in the previous Section the standard (null-hypothesis)
minimum-variance estimator $\fnlnull$ is constructed 
under the null hypothesis, so its variance is strictly minimized
only  when applied to maps with $\fnl =0$
\cite{Creminelli:2006gc}.  In particular, the variance of
$\fnlnull$ is given in Eq.~(\ref{eq:var})
so that when $36 A \fnl^2 l_{\rm max}^2/\ln^2(l_{\rm max})
\gtrsim 1$, the variance scales as the $\ln^{-3}(l_{\rm
max})$, as opposed to $l^{-2}_{\rm max}\ln^{-1}(l_{\rm max})$.  This indicates that when
$\fnl \neq 0$ there may be other estimators with smaller
variances.

For a flat-sky and under the Sachs-Wolfe approximation
Ref.~\cite{Creminelli:2006gc} introduced an improved estimator 
for $\fnl \neq 0$ which has a variance that continues to
decrease as $1/[l_{\rm max}^2 \ln(l_{\rm max})]$ in the high
signal-to-noise limit.  To achieve this scaling they introduced
a realization-dependent normalization, 
\begin{equation}
     \mathcal{N} \equiv \sigma_{\fnl}^2 \sum_{\vec l_1 + \vec
     l_2+\vec l_3=0} \frac{\chi_{\vec l_1} T_{\vec l_2}
     T_{\vec l_3}}{2 C_{l_1} C_{l_2} C_{l_3}}B(l_1,l_2,l_3),
\end{equation}
where 
\begin{eqnarray}
     \chi_{\vec l} &\equiv& \sum_{\vec k} T_{\vec l - \vec k}
     T_{\vec k}.
\end{eqnarray}
By construction $\VEV{\mathcal{N}} = 1$.  They then define a new 
estimator constructed under the non-null hypothesis:
\begin{equation}
\fnlnew \equiv \frac{\fnlnull}{\mathcal{N}}.
\end{equation}
To explore the properties of the PDF of $\fnlnew$, we expand
the normalization as 
$\mathcal{N} \approx \mathcal{N}_0 + \fnl \mathcal{N}_1 + \cdots$ and write
\begin{eqnarray}
\fnlnew &\approx& \frac{{E}_0}{\mathcal{N}_0} + \fnl \frac{{E}_1 \mathcal{N}_0 - 
{E}_0 \mathcal{N}_1}{\mathcal{N}_0^2} + \cdots,\\
&\equiv& \mathcal{E}_0 + \fnl \mathcal{E}_1+ \cdots.
\end{eqnarray}

\begin{figure}[htbp]
\resizebox{!}{6cm}{\includegraphics{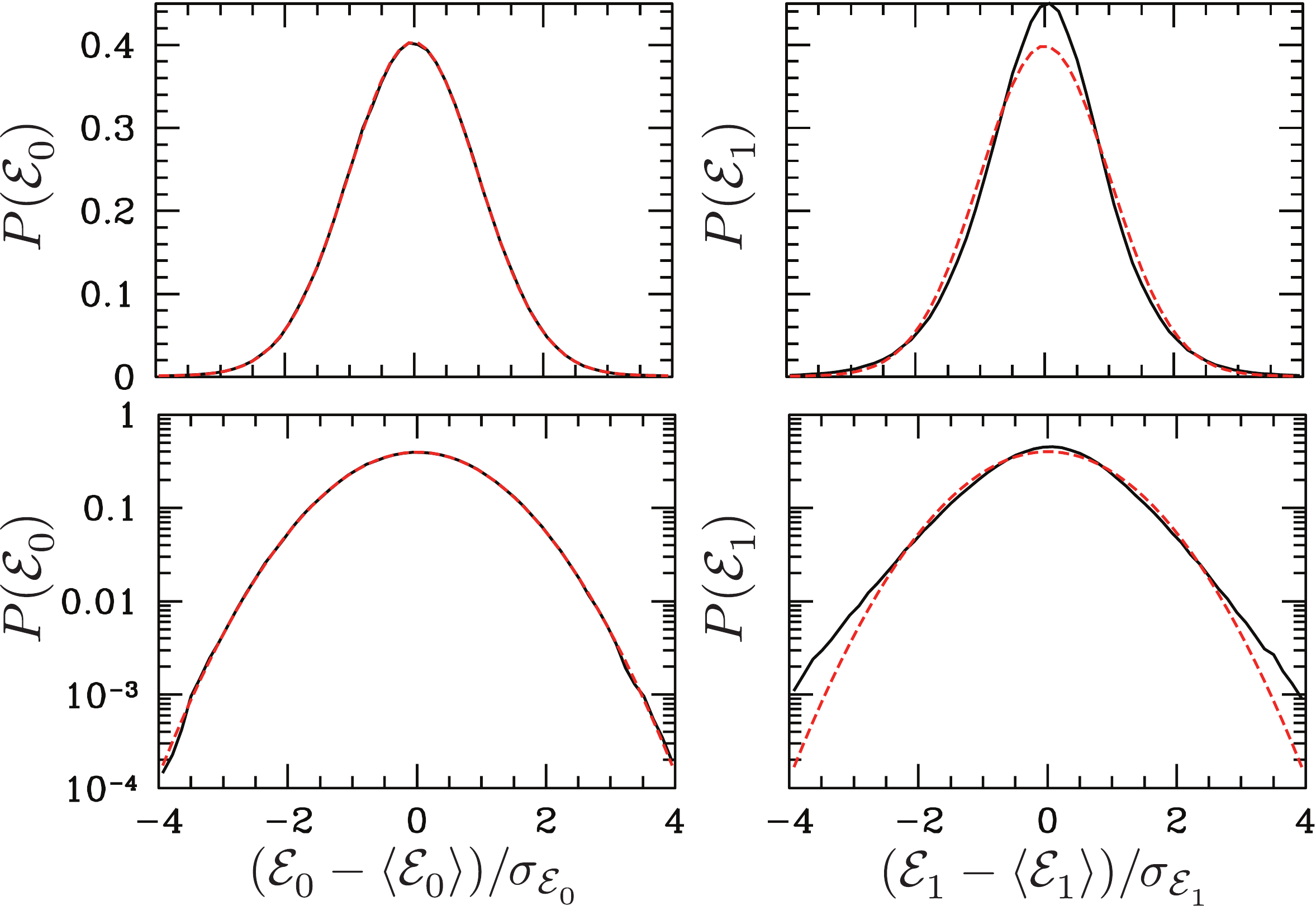}}
\caption{
     The PDF $P(\mathcal{E}_0)$ (left) and $P(\mathcal{E}_1)$ (right)
     for $l_{\rm max} =25$ determined with $10^6$ non-Gaussian 
     realizations.  The top panels show the PDF on a linear scale;
     the bottom panels show the PDF on a log scale.  We have 
     confirmed that the shape of the PDF is unchanged for larger values of $l_{\rm max}$. 
     The PDF of $\mathcal{E}_0$ (left) is well 
     approximated by a Gaussian.  However, the PDF of the 
     first-order correction $\mathcal{E}_1$ (right) has
     significant non-Gaussian
     wings.  This implies that the full PDF of $\fnlnew$ 
     is also non-Gaussian, even if the true value of $\fnl$
     matches that assumed in the construction of the CSZ
     estimator.  Quantitatively, however, the level
     of non-Gaussianty will be small for Planck, as the variance
     of $\mathcal{E}_1$ is $\VEV{(\Delta \mathcal{E}_1)^2}
     \approx 9 \ln^2(l_{\rm max})/(l_{\rm max}^3)$.}
\label{fig:betterPDF}
\end{figure}

In order to determine the shape of $P(\fnlnew)$, we computed
$P(\mathcal{E}_0)$ and $P(\mathcal{E}_1)$ for various values of
$l_{\rm max}$.  We found, as in the $\fnlnull$ case, 
that these PDFs approach asymptotic shapes in the $l_{\rm max} \gg 1$ limit.  We show 
these PDFs in Fig.~\ref{fig:betterPDF} determined by $10^6$ realizations for $l_{\rm max} =25$.
It is clear that $P(\mathcal{E}_0)$ is very well approximated by
a Gaussian, whereas 
$P(\mathcal{E}_1)$ has significant non-Gaussian wings.  As in the 
$P(\fnlnull)$ case, this implies that the level of
non-Gaussianity in $P(\fnlnew)$ is significant only 
when the ratio $\fnl^2 \VEV{(\Delta \mathcal{E}_1)^2}/\VEV{(\Delta
\mathcal{E}_0)^2} \gtrsim 1$.  Our simulations show 
\begin{eqnarray}
     \VEV{(\Delta \mathcal{E}_0)^2} &\approx& \frac{1}{8 A
     l_{\rm max}^2 \ln(l_{\rm max})},\\
     \VEV{(\Delta \mathcal{E}_1)^2} &\approx& 9\frac{\ln^2(l_{\rm
     max})}{l_{\rm max}^3},
\end{eqnarray}
so that the PDF will be significantly non-Gaussian when
\begin{equation}
 \fnl A^{1/2} \gtrsim \frac{1}{3}\left[\frac{l_{\rm max}}{8 \ln(l_{\rm max})}\right]^{1/2}.
\end{equation}
Therefore, for Planck (with $l_{\rm max} = 3000$) $P(\fnlnew;
\fnl,l_{\rm max})$ will be significantly non-Gaussian only if
$\fnl \gtrsim \mathcal{O}(1000)$.  Since this has already been ruled out by
observations \cite{limits,halos}, we conclude that $P(\fnlnew;\fnl,\lmax)$ will be
effectively Gaussian.

\section{Discussion}
\label{sec:discussion}

Here we have argued that the PDF for non-Gaussianity estimators
cannot be assumed to be Gaussian, since the number of triplets
used to construct these estimators may greatly exceed the number
$\Npix$ of measurements.  The 99.7\% confidence-level
interval cannot safely be assumed to be 3 times the 66.5\%
confidence-level interval.  We found, however, that the standard
minimum-variance estimator $\fnlnull$ constructed under
the null hypothesis is well-approximated by a Gaussian distribution
in the $l_{\rm max} \gg 1$ limit if the null hypothesis is
correct (i.e., when applied to purely Gaussian maps).  

We then calculated the same PDF $P(\fnlnull)$
under the hypothesis that the true value of $\fnl$ is non-zero.
We find that the PDF is non-Gaussian in this case, skewed to
large $\fnle$ if $\fnl>0$ and {\it vice versa} for $\fnl<0$.
The PDF for small positive or for negative $\fnle$ is significantly 
smaller for $\fnl>0$ than the Gaussian PDF with the same
variance.  Thus, for example, if the NHMV estimator gives 
$\fnle>0$, it may actually rule out $\fnl=0$
with a smaller statistical significance than would be inferred
assuming a Gaussian distribution of the same variance.  For
Planck (with $l_{\rm max} \simeq 3000$) we find that the
non-Gaussian shape of $P(\fnlnull)$ is significant if $\fnl
\gtrsim \mathcal{O}(10)$.  Thus, the non-Gaussian
shape of the PDF may need to be taken into account, 
even in case of a null result, to assign a precise 99.7\%
confidence-level upper (or lower, for $\fnl<0$) limit to $\fnl$.  We 
also provide, in Eq.~(\ref{eq:fit}), an analytic fit to these PDFs. 

The non-Gaussian shape of $P(\fnlnull)$ when $\fnl \neq 0$ is accompanied by 
a variance that decreases only logarithmically with increasing $l_{\rm max}$. 
Because of this, Ref.~\cite{Creminelli:2006gc} constructed an
improved estimator under the $\fnl \neq 0$ hypothesis with a
variance that saturates the Cramer-Rao bound 
and continues to decrease as $1/[l_{\rm max}^2\log(l_{\rm max})]$.   We
found that for observationally allowed values of $\fnl$ this
improved estimator has a PDF that is well approximated by a
Gaussian shape.  However, this estimator has only been defined under the 
Sachs-Wolfe limit and it is not immediately clear how it should be 
generalized to be applied to actual data.  An alternative, Bayesian, approach 
to measuring $\fnl$ which also saturates the Cramer-Rao bound 
in the presence of $\fnl \neq 0$ is presented in Ref.~\cite{Elsner:2010hb}.

We have restricted our attention to the bispectrum in the local
model, but the PDF must be similarly determined for the
non-Gaussianity parameter for bispectra with other shape
dependences; e.g., the equilateral model \cite{dvali,Equil} or
that which arises with self-ordering scalar fields
\cite{Figueroa:2010zx}.  It should also be interesting to
explore the PDF for maximum-likelihood, rather than
quadratic, estimators (see, e.g.,
Ref.~\cite{Creminelli:2006gc}).  Ultimately, a variety of
experimental effects and more precise power spectra and
bispectra, rather than the Sachs-Wolfe-limit quantities used
here, will need to be included in interpreting the results of
realistic experiments.

There is also interest in using higher-order correlation functions 
to measure $\fnl$ from CMB maps.
Our arguments should apply also to these higher-order
correlation functions, like the trispectrum, etc.  
For example, the estimator for the amplitude of the $n$-point
correlation function (e.g., $n=3$ for the bispectrum, $n=4$ for the
trispectrum, etc.), will be constructed from $\sim
\Npix^{(n-1)}/n!$ combinations of $n$ pixels, and this number of
combinations scales even more rapidly with $\Npix$ than that for
the bispectrum.  Thus, although the signal-to-noise scales
more rapidly with $\Npix$ for these higher-order correlation
functions than that for the bispectrum
\cite{Kogo:2006kh,Smidt:2010sv,Kamionkowski:2010me}, concerns about the PDF for
these estimators should be even  more serious than for the
bispectrum.  It will thus be necessary to understand the PDF for
these higher-order estimators to confidently forecast the
statistical signficance of measurements \cite{inprep}.

\begin{acknowledgments}
We thank D.~Babich, C.~Hirata, and I.~Wehus for useful
discussions. TLS is supported by the 
Berkeley Center of Cosmological Physics. 
MK thanks the support of the Miller Institute and
the hospitality of the Department of Physics at the University of
California, where part of this work was completed.
MK was supported at Caltech by DoE DE-FG03-92-ER40701,
NASA NNX10AD04G, and the Gordon and Betty Moore Foundation.
BDW was supported by NASA/JPL subcontract 1413479,
and NSF grants AST 07-08849,  AST 09-08693 ARRA, and AST 09-08902 during this work.
\end{acknowledgments}

\begin{appendix}

\section{Computing non-Gaussianity estimators using FFTs}

We are interested in using Monte Carlo simulations to determine
the shape of the PDF of $\hnlb$ as a function of the fiducial
choice of $\fnl$ and the number $\Npix$ of pixels measured in a
given observation.  Applying the estimator in
Eq.~(\ref{eqn:biestimator}) to the local-model bispectrum
[Eq.~(\ref{eq:localModBi})] it can be rewritten
\begin{equation}
     \hnlb = \sigma_{\fnl}^{2} \sum_{|\vec l _1+ \vec l_2+ \vec l_3|=0}
     \frac{ T_{\vec l_1} T_{\vec l_2} T_{\vec l_3}}{\Omega^2 C_{l_3}}.
     \label{eqn:localbiestimator}
\end{equation}
The estimator in Eq.~(\ref{eqn:localbiestimator}) 
takes $\Npix^2$ 
operations to evaluate.  Since current CMB observations 
have $\Npix \sim 10^6$ this estimator would take a prohibitively long time to evaluate for a 
significant number of realizations, especially since we are interested in probing the shape of the 
PDF far into the tail of the distribution ($\sim 3-4 \sigma$).  

As discussed at length in Ref.~\cite{Komatsu:2003iq} this is even more of a problem when 
measuring non-Gaussianity on the full sky where the number of operations  
scales as $\Npix^{5/2}$.  In order to make the problem tractable 
Ref.~\cite{Komatsu:2003iq} rewrites $\hnlb$ in terms of real-space quantities 
reducing the number of operations to $\Npix^{3/2}$.  

We can do the same for $\hnlb$ in the flat-sky approximation.  Noting that 
\begin{equation}
     \delta_{\vec l_1+ \vec l_2 + \vec l_3, 0} = \int\frac{d^2 \theta}{\Omega} e^{i
     \vec{\theta} \cdot (\vec l_1 + \vec l_2+ \vec l_3)}, 
\end{equation}
and writing 
\begin{eqnarray}
 A(\vec \theta)  &\equiv&
     \frac{1}{\Omega} \sum_{\vec l} e^{i\vec l\cdot
     \vec \theta}T_{\vec l},\label{eq:T}\\
     B(\vec \theta)  &\equiv&
     \frac{1}{\Omega} \sum_{\vec l} e^{i\vec l\cdot
     \vec \theta}\frac{T_{\vec l}}{C_l}, \label{eq:chi}
\end{eqnarray}
$\hnlb$ can be written 
\begin{equation}
     \hnlb = \Omega \sigma_{\fnl}^2 \int \frac{d^2 \theta}{\Omega}\
     A^2(\vec{\theta})B(\vec{\theta}).
\label{eq:intfnle}
\end{equation}
Next, in order to compute the integral in Eq.~(\ref{eq:intfnle})
we use the Nyquist sampling theorem and the fact that both
$A(\vec{\theta})$ and $B(\vec{\theta})$ have finite Fourier
spectra (truncated at a maximum frequency $l_{\rm max}$).  This
allows us to rewrite the integral as a discrete sum 
\begin{eqnarray}
\hnlb &=& \frac{\Omega \sigma_{\fnl}^2}{N^2} \sum_{i=1}^{N}\sum_{j=1}^{N}
A^2\left(2 \pi \frac{i-1}{N},2 \pi \frac{j-1}{N}\right)\nonumber \\
&\times& B\left(2 \pi \frac{i-1}{N},2 \pi \frac{j-1}{N}\right),
\end{eqnarray}
where $N \equiv 2(2 l_{\rm max}+1)$.

Since Eqs.~(\ref{eq:T}) and (\ref{eq:chi}) are discrete inverse Fourier transforms we can use a 
fast Fourier transform (FFT) 
algorithm so that the number of operations scale as $\Npix \ln(\Npix)$. 

We can use the same computational trick when evaluating the
non-Gaussian contribution for each realization by also employing
a forward FFT in order to compute the convolution in
Eq.~(\ref{eq:nonGauss}).

\section{Analytic calculation of $\VEV{(\Delta{E}_1)^2}$}

In order to verify that our simulations are correct we performed
an analytic calculation of the variance of ${E}_1$
[Eq.~(\ref{eq:E1})] defined by
\begin{equation}
     {E}_1 = \sigma_{\fnl}^2 \sum_{\vec l_1+ \vec l_2+ \vec l_3=0} 
     \frac{\delta t^2_{\vec l_1}t_{\vec l_2}t_{\vec l_3}}{2
     C_{l_1} C_{l_2} C_{l_3} }B(l_1,l_2,l_3).
\end{equation}
A straightforward but tedious calculation shows that the variance is given by
\begin{equation}
     \VEV{(\Delta  E_1)^2} = 9\sigma_{\fnl}^4 (A_1+8 A_2 + A_3 +4 A_4),
\end{equation}
where 
\begin{eqnarray}
     A_1 &\equiv& \sum_{\{\vec l\}, \{\vec k\}}
     \frac{B(l)}{C_{l_1}}\frac{B(k)}{C_{k_1}} \delta_{\vec l_1 +
     \vec k_1,0} ,\\ 
     A_2 &\equiv&  \sum_{\{\vec l\}, \{\vec k\}}
     \frac{B(l)}{C_{l_1}}\frac{B(k)}{C_{k_1}} \delta_{\vec l_3 +
     \vec k_3,0},\\
     A_3 &\equiv&  \sum_{\{\vec  l\}} \frac{B(l)^2}{C_{l_1}^2
     C_{l_2} C_{l_3}} \sum_{|\vec m| = 1}^{l_{\rm max}} C_{|\vec
     l_1 -\vec m|}C_m ,\\
     A_4 &\equiv& \sum_{\{\vec l\}, \{\vec k\}} \frac{B(l)
     B(k)}{C_{l_1} C_{k_1} C_{k_3}} C_{|\vec l_1 + \vec
     k_2|}\delta_{l_3+k_3,0},
\end{eqnarray}
where $\{ \vec l\}$ indicates the sum is over $|\vec l_1+\vec
l_2+\vec l_3|=0$ and $B(l) \equiv B(l_1,l_2,l_3)$.  Computing
these terms as a function of $l_{\rm max}$ we find that the
variance is well-fit by the function 
\begin{equation}
     \VEV{(\Delta {E}_1)^2} = \frac{14.0\ l_{\rm
     max}^{0.433}}{\ln^{5.1}(l_{\rm max})}.
\end{equation}
In Fig.~\ref{fig:varE1} we show how that this analytic
calculation of the $\VEV{(\Delta {E}_1)^2}$ is reproduced
by the results of the Monte Carlo simulations.  

\end{appendix}

\end{document}